# Magnetic Structure and dielectric properties of antiferromagnetic FeCrO$_3$

Rajesh kumar R[1], C. Dhanasekhar[3,4], N. Vijay Prakash Chaudhary[2], A. Das[5,6] and A. Venimadhav[1,2]

[1]School of Nanoscience and Technology, Indian Institute of Technology, Kharagpur-721302, India
[2]Cryogenic Engineering centre, Indian Institute of Technology, Kharagpur-721302, India
[3]Department of Physics, Indian Institute of Technology, Kharagpur-721302, India
[4]Department of Physics, Indian Institute of Technology, Mumbai- 40076, India
[5]Solid State Physics Division, Bhabha Atomic Research Centre, Mumbai 400085, India
[6]Homi Bhabha National institute, Anushaktinagar, Mumbai-400094, India
venimadhav@cryo.iitkgp.ac.in
r.g.rajeshkumarr@gmail.com

**Abstract**

We report the enhancement of Néel temperature of Cr$_2$O$_3$ by replacing 50% of Cr by Fe prepared by sol-gel method. The structural analysis by neutron diffraction has revealed that FeCrO$_3$ belongs to a corundum structure (R-3c space group) with an antiferromagnetic spin structure having collinear spins along *a*-axis with propagation vector k=0; the high-temperature magnetisation study indicated a Néel temperature of 560 K. The enhancement in Néel temperature has been attributed to the strong orbital hybridization that leads to change in nearest neighbor bond angle and bond distances. Impedance spectroscopy has revealed conduction mechanism at low temperature is due to the polaron hopping while extrinsic contributions from the Maxwell-Wagner dominant at high temperatures.

Corresponding authors

venimadhav@cryo.iitkgp.ac.in

r.g.rajeshkumarr@gmail.com

**Introduction**

Antiferromagnets are promising for spintronics due to their non-volatility, no stray fields and robustness against external magnetic fields. Insulating antiferromagnets are special and attractive for magnonic effects, caloritrons and optical excitation of spin waves [1, 2]. Insulating antiferromagnets are mostly oxides and in special cases display magnetoelectric effect (ME). Corundum $Fe_2O_3$ and $Cr_2O_3$ are attractive from the above considerations. The linear ME effect was first observed in chromium oxide, $Cr_2O_3$ from the pioneering works of Dzyaloshinskii [3] and Astrov [4]. In recent years, $Cr_2O_3$ has become one of the important ME materials and studied in bulk, single crystals and epitaxial films [5-10]. $Cr_2O_3$ crystallizes in corundum structure with R-3c space group and exhibits antiferromagnetic (AFM) ordering with Néel temperature ($T_N$) of 307 K; having -3′m′ magnetic symmetry; it breaks the space-inversion under magnetic field that allows the linear ME effect in this material [11, 12]. $Cr_2O_3$ has emerged as a key material for realizing low energy consuming spintronics devices based on ME memory and storage [13, 14].

The $T_N$ of $Cr_2O_3$ falls below room temperature in thin heterostructures which impedes its practical usage [15, 16]. There have been several attempts to overcome this problem by hydrostatic pressure and doping with several elements like nickel (Ni), cobalt (Co), iron (Fe), titanium (Ti), manganese (Mn), nitrogen (N) and boron (B) [17-21]. Among these, doping with boron has shown to increase Néel temperature from both theoretical and experimental studies [17, 19]. However, having high $T_N$ of $Fe_2O_3$ is expected to enhance the $T_N$ of $Cr_2O_3$ with Fe substitution. Although $Cr_2O_3$ and $Fe_2O_3$ are crystallographically identical, their magnetic structures are different. In $Cr_2O_3$, Cr -O- Cr ions are AFM coupled along the interlayers, while, in α-$Fe_2O_3$ ($T_N$ ~ 950 K), Fe -O- Fe ions have a ferromagnetic coupling (FM) within the layer and AF to adjacent layers along [0001] direction and antiferromagnetic spin configuration of $Fe_2O_3$ changes its direction from base plane to c-axis at Morin transition (~260 K) [22]. Though there are first principle studies on electronic and magnetic structure of Fe doping in $Cr_2O_3$ [23, 24], there are only fewer experimental investigations [25, 26] and no reports on the dielectric properties.

In this study, we examined the effect of substitution of 50% of Fe in Cr site on $T_N$ of $Cr_2O_3$. A detailed structural and magnetic investigation has been carried on $FeCrO_3$ sample prepared by sol-gel method. The high-temperature dc magnetisation measurement has indicated the Néel temperature of 560 K in the material well above room temperature. The dielectric study suggested the conduction mechanism is due to the polaron hopping conductivity and extrinsic contribution from the Maxwell-Wagner effect at high temperature.

**Experimental details**

The $FeCrO_3$ (FECO) has been prepared by a sol-gel synthesis method. A stoichiometric mixture of the precursors $Cr(NO_3)_2 \cdot 6H_2O$ and $Fe(NO_3)_2 \cdot 6H_2O$ were mixed in citric acid ($C_6H_8O_7 \cdot H_2O$) and heated overnight at 80°C to obtained the gel. The gel was washed several times with ethanol and distilled water to remove the unreacted precursors and dried in an oven. The oven dried materials were mixed and homogenised through grinding for 30 min with an agate mortar and pestle. The powder was pressed into cylindrical pellets and sintered at 1300°C for 24 h and cooled slowly to room temperature. Phase purity and grain size were studied by X-ray diffraction (XRD) technique with Philips panalytical diffractometer with Cu Kα radiation (λ= 1.5405 Å). The neutron diffraction measurements of $FeCrO_3$ samples were carried out on the PD2 powder neutron diffractometer (λ=1.2443 Å) at Dhruva reactor, Bhabha Atomic Research centre, Mumbai, India. The powder samples of approximately 5g were packed in a cylindrical vanadium container and attached to the cold finger of a closed cycle helium refrigerator. Rietveld refinement of the neutron diffraction pattern were carried out using FULLPROF program. Magnetic measurements were carried out in a commercial VSM SQUID magnetometer with oven module (Quantum Design, USA). Optical absorption measurements were performed at room temperature using a Cary 5000 spectrophotometer in diffusion reflectance mode. Dielectric measurements were carried out in homemade setup using an Agilent HP4294A impedance analyser. High purity silver paste electrode contacts were applied on both faces of the pellet to form a parallel capacitance geometry.

**Results and discussion**

The crystal and magnetic structure of the materials is studied by analyzing the temperature dependent neutron diffraction patterns recorded at selected temperature in the range of 7 K ≤T ≤ 300 K. Fig 1 (a-c) shows the diffraction pattern at 300 K, 150 K and 7 K for $FeCrO_3$ and the Rietveld refinement shows no impurity phase present in the sample. $FeCrO_3$ shows corundum crystal structure with R-3c space group, and is shown in Fig. 2. In the corundum structure $Cr^{3+}/Fe^{3+}$ occupy the 12c positions with the coordination (0, 0, z) and surrounded by 6 oxygen ions forming the octahedral whereas the oxygen occupy 18e positions with (x, 0, z). The results of the refinement of the corundum phase at 300 K and 7 K are summarized in

Table 1 respectively. An increase in the unit cell volume was observed with Fe doping is evident due to the higher ionic radii of $Fe^{3+}$ (0.645 Å) ion as compared to the $Cr^{3+}$ (0.615 Å) ion.

The diffraction patterns at different temperatures shows the presence of both fundamental and superlattice magnetic reflections indicating the magnetic nature of the sample. The intensity of the magnetic reflections (003) and (101) are indexed in P-1 space group exhibits enhancement in the intensity with the decrease of temperature. The (003) and (101) are purely magnetic and are absent in the $Cr_2O_3$ [18, 19].It indicates that the magnetic structure of $FeCrO_3$ is different from that of the parent, in spite having same crystal structures. The magnetic structure of this compound is different from the samples with (1-x) $Cr_2O_3$-x $Fe_2O_3$ with 0.1 < X < 0.4, where a non collinear magnetic structure is observed, however it is closer to $Fe_2O_3$ above Morin transition [27]. Fig. 2 also shows the magnetic structure of $FeCrO_3$ in the corundum structure with propagation vector, k=0, implying identical chemical and magnetic cell. The AFM spin arrangement of Fe/Cr ions along the *a*-axis in the temperature range between 300 K and 7 K and confirms the absence of Morin transition similar to that observed in $Fe_2O_3$ [27, 28].

**High temperature magnetization studies**

From the structural data, it is evident that the sample is magnetic without having a transition temperature below 300 K. It implies that the transition is above the room temperature and accordingly the magnetization of the sample at high temperature has been studied. The temperature dependent magnetization zero field cooled (ZFC) and field cooled warming (FCW) was recorded in the temperature range from 300-750 K. The variation of ZFC (closed circles) and FCW (open circles) magnetisation curves with temperature under an applied magnetic field of 100 Oe is shown in Fig. 3(a). ZFC curve is shown in the insert of figure 3(a) indicates that the Néel temperature $T_N$ = 560 K for $FeCrO_3$ and the irreversibility between the ZFC and FCC curves persist even at high temperatures indicating the high anisotropy in the material.

The linear variation of $1/\chi_{mol}$ with respect to temperature indicates the realization of Curie-Weiss law. According to the Curie-Weiss law temperature dependence of the molar magnetic susceptibility can be described by $\chi_{mol} = C_M/T-\theta$, where $C_M$ is a molar Curie constant and $\theta$ is the Curie Weiss temperature. Fitting the magnetization with the Curie-Weiss law gives Curie Weiss temperature ~ -1062 K and the obtained spin only moment of ~ 8.32 $\mu_B$/f.u for

FeCrO$_3$, this value is more than that of Cr$_2$O$_3$ (5.47 µ$_B$/f.u.) and closely matches with Fe$_2$O$_3$ (8.37 µ$_B$/f.u.). However, the obtained effective moment of FeCrO$_3$ is more than the theoretical value of 7.07 µ$_B$/f.u. Such a mismatch between the theoretical and experimental values could be accounted from orbital contribution of Cr$^{3+}$ or exchange randomness or combination of the both [27, 29]. Isothermal magnetic hysteresis loop measurements carried out in the temperature range of 5 K - 400 K are shown in Fig. 3 (b) exhibits antiferromagnetic behaviour throughout the temperature.

The magnetic interaction in the corundum structures can be understood from the Goodenough-kanamori (GK) rules. Accordingly, the first nearest neighbour J1 and second nearest neighbour J2 favours the aniferromagnetic coupling and the magnitude of J1 and J2 are larger than the other J3, J4 and J5 interactions as shown in Fig. 2. In corundum type Cr$_2$O$_3$, the Cr-O-Cr bond angles corresponding to J1 (82.5°) and J2 (93.1°) are close to 90°, whereas the other nearest neighboring interactions J3 (121.1°) and J4 (133.2°) are close to 135°. In FeCrO$_3$, substituting Fe in Cr site changes the bond angles associated with the J1 (84.73°), J2 (93.81°), J3 (119.84°) and J4 (132.45°) exchanges interactions. These changes originate from hybridization between of Cr$^{3+}$ is (t$_{2g}^3$, eg$^0$) and Fe$^{3+}$ is (t$_{2g}^5$, eg$^2$), Cr provides the charge transfer from Cr t$_{2g}$ and O 2p states to Fe t$_{2g}$* this hybridization gives rise to the σ-type super exchange [30, 31] in the FeCrO$_3$. The J1 and J2 with bond angle close to the 90° can be attributed to the kinetic exchange induced by direct hybridization between the t$_{2g}$ orbitals, further, the superexchange via oxygen is weak due to the face and edge sharing octahedral; and this kinetic exchange increase the bond angle of both J1 and J2. In Cr$_2$O$_3$, J3 and J4 with larger bond angles close to 135° have larger distances as the direct superexchange via oxygen dominates over kinetic exchange due to the absence of eg electron and this interaction is weak. While this superexchange (antiferromagnetic) can dominate in FeCrO$_3$ as Fe contains e$_g$ electrons. The J5 coupling may be neglected as it is mediated through more than two oxygen atoms.

The hybridization in FeCrO$_3$ is supported by the band gap analysis using diffusion reflectance mode of UV-Visible spectroscopy using the general equation α ($hv$) ≈ B ($hv$ − E$_g$)$^n$; where E$_g$ is the band gap (eV), $h$ is the Planck's constant, B is the absorption constant, $v$ is the light frequency and α is the extinction coefficient. The corresponding coefficient (n) associated with an electronic transition as follows and it is 1/2 for indirect band gap and 2 for direct band gap. As shown in Fig. 4, FeCrO$_3$ exhibits a direct band gap of 1.67 eV which is lower than that of parent compounds, Fe$_2$O$_3$ (~ 2.2 eV) and Cr$_2$O$_3$ (~3.2 eV); this band gap is previous theoretical and experimental results [24, 31, 32]. Narrowing of band gap is attributed to the transition from occupied Cr $t_{2g}$ and O $2_p$ states to empty Fe $t_{2g}^*$ states [30-32] which helps in strong hybridization and lowering of band gap in $FeCrO_3$.

**Impedance spectroscopy**

Temperature dependent impedance spectroscopy analysis was carried out to understand the conduction mechanism in the material. Complex impedance analysis is a powerful technique, especially for examining the conduction mechanisms as well as the separation of various contributions to the dielectric behavior such as from grains, grain boundaries, space charge or electrodes.

Fig. 5(a-b) shows the temperature dependent variation in real and imaginary parts of the dielectric permittivity ($\varepsilon'$ and $\varepsilon''$). The imaginary part of the dielectric $\varepsilon''$, exhibits relaxation in between 125 K - 250 K, these relaxation peaks are associated with the real part of the $\varepsilon'$. The relaxation peak is found to move to the higher temperature side with increase in frequency. In the insert of Fig. 5(a) the tan$\delta$ and $\varepsilon''$ with respect to temperature is plotted for 10 kHz and a large temperature separation of peaks between tan$\delta$ and $\varepsilon''$ peaks rules out the existence of permanent dipoles in $FeCrO_3$ sample. In fact, the appearance of relaxor like behavior in an electrically heterogeneous system can also occur due to the crossover in the relaxation time associated with the grains and grain boundaries in accordance with the Maxwell–Wagner interfacial model [30, 33]

We have analyzed the relaxation mechanism by the thermally activated Arrhenius behaviour given

$$\tau = \tau_o \exp(\frac{E_a}{K_B T}) \qquad - (1)$$

Where $E_a$ is the activation energy required for the relaxation process and $\tau_0$ is the pre-exponential factor. The plot of ln$\tau$ versus 1/T is shown in the insert of Fig. 5(b), where the solid lines indicate the fit to equation (1), it is seen that the fit deviates in the high temperature side, therefore two fits have been are done for different temperature ranges. The activation energy and relation time towards low temperature side of the fit are 0.15 eV and 6.6945 x $10^{-11}$s, whereas towards high temperature side are 0.21 eV and 5.82108 x $10^{-9}$s, respectively.

To capture the essential physical process of the system representing the data is represented by the electrical modulus (M"). Fig. 6(a) shows the imaginary part of M" as a function of temperature, exhibits two relaxation peaks, one at high temperature and the other at low temperature for all frequencies. It can be noticed that the high temperature exhibit much stronger dispersion compared to the low temperature relaxation. The lower frequency side of the frequency spectrum can be assigned to the grain contribution while higher temperature side to grain boundary. The relaxation mechanism at grain and grain boundaries are analyzed using the thermally activated Arrhenius mechanism, and represented in Fig 6 (b). The grain relaxation show activation energy of 0.17 eV with a relaxation time of $2.39 \times 10^{-10}$ S; the value of the activation energy is comparable to the polaronic formation energy [34, 35]. While the grain boundary (GB) activation energy of 0.49 eV and relaxation time of $2.8452 \times 10^{-12}$ S can be attribute to the movement of oxygen vacancies across the GB [36-38].

**Conclusion**

In conclusion, 50% Fe substituted $Cr_2O_3$ prepared by sol-gel method crystalizes in corundum structure. The structural and magnetic properties studied by neutron diffraction showed the AFM spin arrangement of Fe/Cr ions along *a*-axis. It is significant to note that the Néel temperature enhances to 560 K in Fe substituted sample; a strong orbital hybridization of Fe-O-Cr leads to change in nearest neighbor bond angle and bond distances. The temperature dependent dielectric properties shows polaron transport near room temperature. $FeCrO_3$ with high $T_N$ can be very attractive for antiferromagnetic spintronics.

**Acknowledgements**

The authors from IIT Kharagpur acknowledge DST (India) for FIST project and IIT Kharagpur (INDIA) for funding the VSM SQUID magnetometer and SGBSI-CHI-138. Author Rajesh kumar acknowledges MHRD, Delhi for senior research fellowship (SRF).

Figure captions

Fig. 1. (a-c) Neutron diffraction pattern for FeCrO$_3$ at 7 K, 150 K, and 300 K. The markers from top to bottom indicate nuclear and magnetic positions respectively. (d) Configuration of Fe/Cr spins in the corundum crystal.

Fig. 2. (a) shows Crystal structure of FeCrO$_3$ and (b) shows spin structure constructed from neutron study.

Fig. 3. (a) Temperature dependent M-T curves of FeCrO$_3$ at 100 Oe and ZFC curve is show in the insert of the figure. (b) Isothermal hysteresis loop from 5 K to 400 K.

Fig. 4. Band gap of $FeCrO_3$ obtained by fitting to equation $\alpha(h\nu) \approx B(h\nu - E_g)^n$.

Fig. 5. Temperature variation of (a) the real part and $\varepsilon''$ and $\tan\delta$ is shown in the insert (b) $\varepsilon''$ variation with temperature and inset shows plot of $\ln\tau$ versus $1/T$, the solid line indicates the Arrhenius fit.

Fig 6. The temperature variation of (a) imaginary part of the electrical modulus ($M''$). (b) Plots of $\ln\tau$ versus $1/T$ for the grain and grain boundaries peaks and solid lines indicate the fit to equation.

Table

Table 1. Lattice parameters, nearest neighbor bond length and bond angles as obtained from Rietveld refinement of Neutron Diffraction data of FeCrO3 at 7 K and 300 K. The atomic Cr/Fe (0, 0, z) and oxygen (x, 0, z) sites in the R-3c space group.

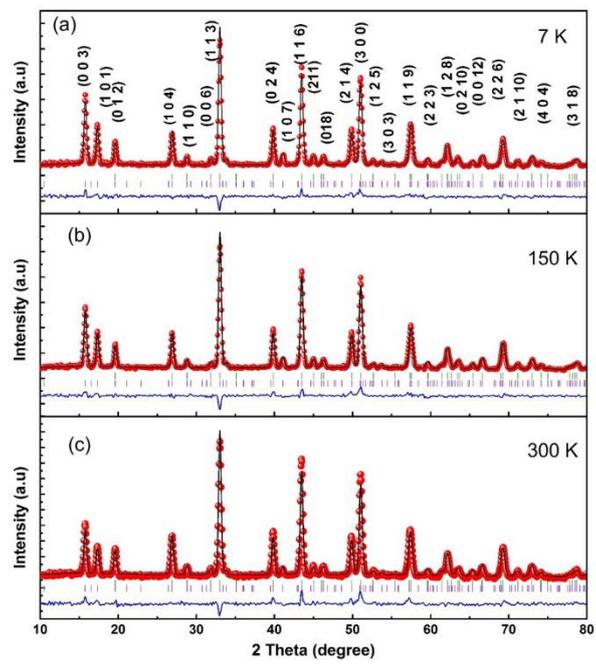

Figures 1

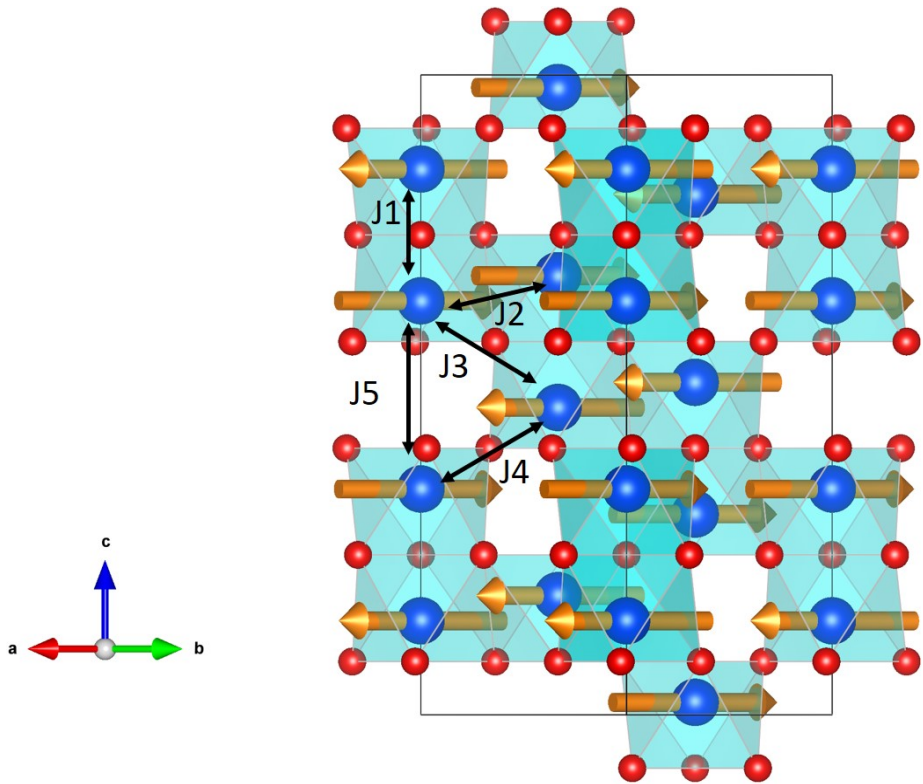

Figure 2

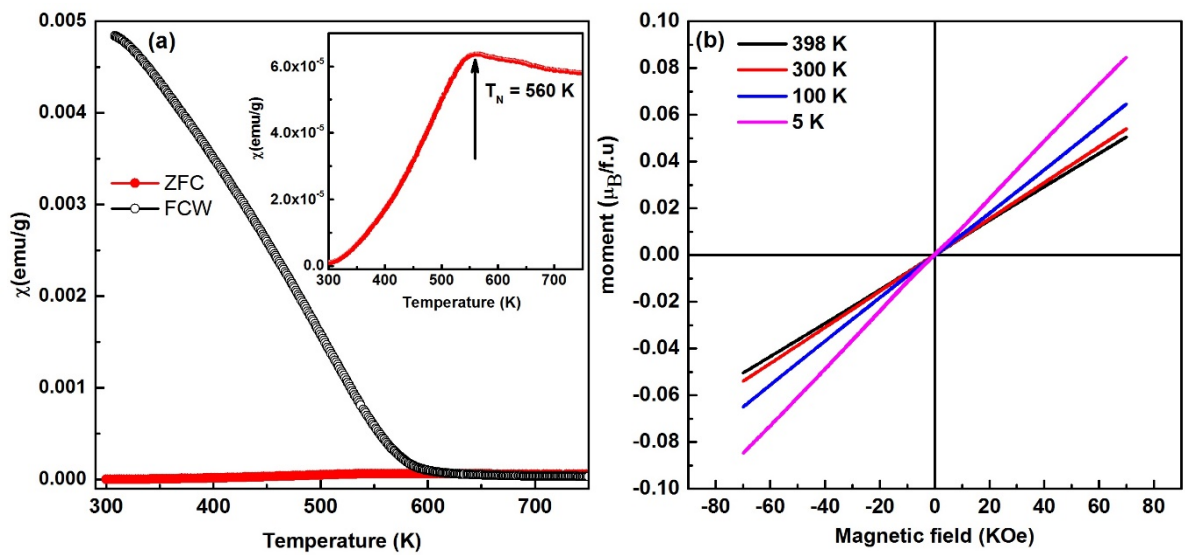

Figure 3

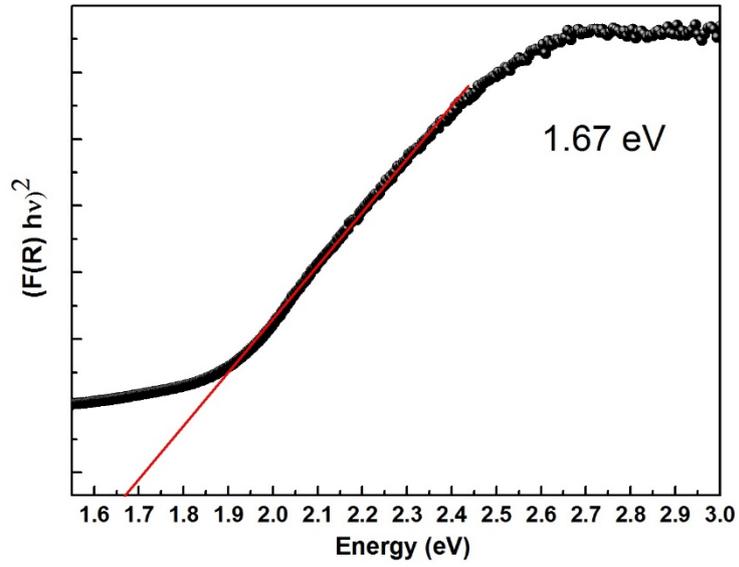

Figure 4

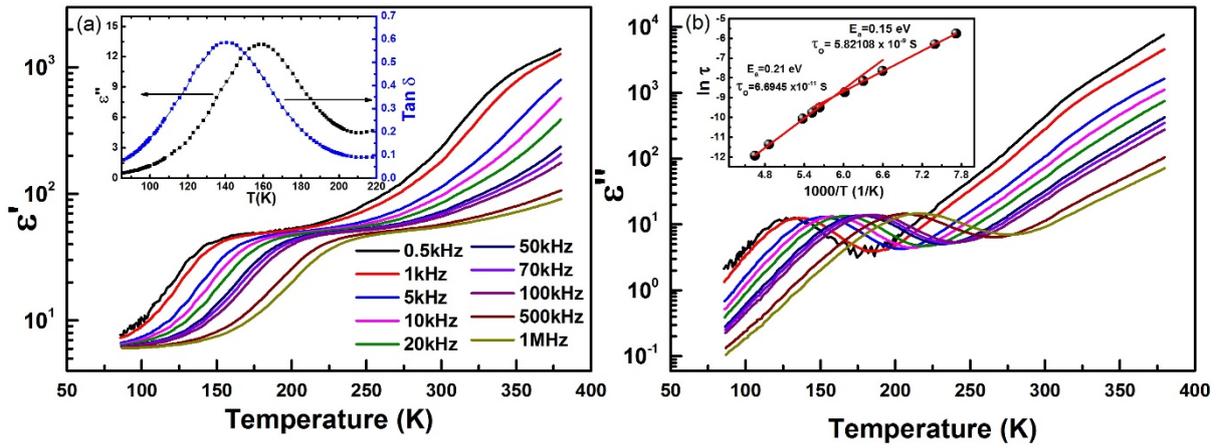

Figure 5

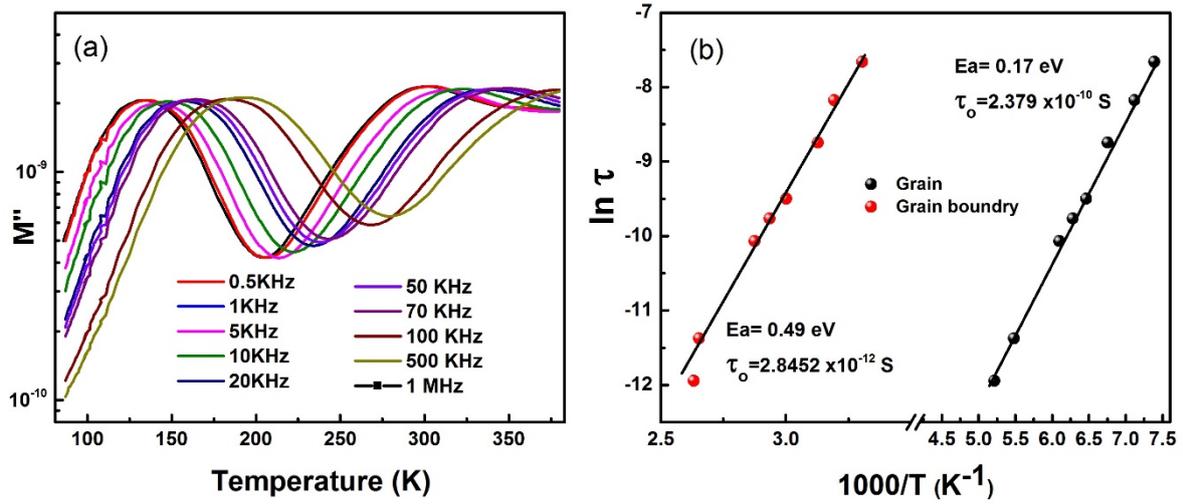

Figure 6

Table 1

|  | 7 K | 300K |
|---|---|---|
| a (Å) | 5.00252(18) | 5.00038(29) |
| c (Å) | 13.5943(76) | 13.6071(11) |
| V(Å$^{-3}$) | 294.6251(13) | 294.6513(16) |
| Nearest Neighbor distances |  |  |
| J1 | 2.85917(17) | 2.8540(3) |
| J2 | 2.94851(8) | 2.94584(13) Å |
| J3 | 3.67088(11) Å | 3.67123(16) Å |
| J4 | 3.33737(9) Å | 3.34106(13) Å |
| J5 | 3.9380(3) Å | 3.9496(4) Å |

| Bond angle between nearest neighbors | | |
|---|---|---|
| J1(deg) | 85.04(7) | 84.58(12) |
| J2 (deg) | 93.924(14) | 93.76(3) |
| J3 (deg) | 132.06(3) | 120.47(11) |
| J4 (deg) | 119.88(6) | 131.98(5) |
| $\chi^2$ | 5.24 | 8.33 |
| $R_{Bragg}$ (%) | 4.64 | 7.43 |